\documentclass[final,3p,times]{elsarticle}

\usepackage{graphicx}
\usepackage{amsmath,mathtools}
\usepackage{color,physics}
\usepackage{amssymb}
\usepackage{amsmath}
\usepackage{graphicx}
\usepackage{amsmath,mathtools}
\usepackage{color,ulem,bm}
\usepackage{amsthm}

\usepackage[hidelinks]{hyperref}

\begin{document}

\begin{frontmatter}

\title{Fischer Information of a Nonequilibrium  Anharmonic Donor-Acceptor Rectifier}

\author{Bitap Raj Thakuria}
\author{Trishna Kalita       }
\author{Javed Akhtar}
\author{
Himangshu Prabal Goswami\corref{cor1}}
\ead{hpg@gauhati.ac.in}
\cortext[cor1]{Corresponding author}

\address{QuAInT Research Group,Department of Chemistry, Gauhati University, GNB Nagar, Jalukbari, Guwahati 781014, Assam, India},
            
\date{\today}
\begin{abstract} We investigate a nonequilibrium donor-acceptor quantum rectifier system coupled to an anharmonic vibrational mode, treating the vibrational dynamics both as a two-level system and as multilevel system. The time-dependent Fischer information is then calculated by deriving a quantum master equation for the reduced system dynamics. We estimate some key rectifier parameters, the donor energy, the acceptor energy, and the vibrational frequency. We report that there is an optimal time for estimating the donor and acceptor energy. However, the anharmonic mode can be estimated better only in the steadystate.  The acceptor energy is found to be most precisely estimable, especially under strong coupling and high bias. Donor energy shows limited sensitivity, while vibrational frequency estimation benefits from low temperatures. This work offers a theoretical foundation for enhancing parameter estimation in nanoscale quantum devices, guiding future sensing and metrological applications in quantronic systems.
\end{abstract}

\begin{keyword}
Quantum Metrology \sep Quantum Junction \sep Nonequilibrium Quantum Systems
\end{keyword}

\end{frontmatter}

\section{Introduction }
Quantum parameter estimation offers a  framework for assessing the sensitivity of open quantum systems to variations in system or environmental parameters~\cite{paris2009,PhysRevA.64.042105,Liu_2020}. In nanoscale transport setups, such as molecular junctions and donor–acceptor complexes, these parameters often include site energies, inter-site couplings, or couplings to vibrational degrees of freedom\cite{Thoss10.1063/1.5003306, Menzeldoi:10.1021/acs.jpclett.9b02408}. Estimating such parameters from steady-state density matrices has become particularly relevant as experimental techniques now allow precision measurements in the low-dimensional, nonequilibrium regime~\cite{Thoss10.1063/1.5003306,Menzeldoi:10.1021/acs.jpclett.9b02408,pekola2015,bruch2016}. It offers a rigorous and non-perturbative measure of how well system parameters can be extracted from quantum state tomography or from observables related to the electronic populations
 In systems where quantum coherence, dissipation, and thermal fluctuations coexist, the steady-state density matrix encodes information about external parameters through its dependence on system-bath interactions and internal couplings. The quantum Fisher information (QFI) provides a metric for quantifying this information content and sets a bound on the precision of unbiased parameter estimation. In contrast to equilibrium systems, nonequilibrium steady states (NESS) offer enhanced sensitivity due to population and coherence imbalances maintained by external drives, leading to richer estimation landscapes \cite{Seah2019, Mancino2018}.
In this context, the quantum Fisher information (QFI) quantifies the maximum achievable precision for estimating a parameter embedded in the quantum state of a system, constrained only by quantum mechanics and independent of the particular measurement strategy~\cite{helstrom1976,PRXQuantum.2.020308}.

This work focuses on a donor–acceptor (DA) system interacting with a localized vibrational mode that is treated as an anharmonic, few-level system\cite{Agarwalla2015}. The vibrational mode couples to the electronic degrees of freedom either by modulating the tunneling amplitude between donor and acceptor or by affecting site energies through polaronic shifts. Importantly, this vibrational degree of freedom remains localized and discrete in character, distinguishing the model from setups with extended phonon baths. The inclusion of vibrational anharmonicity modifies the  open system dynamics and alters the structure of steady states, introducing nonlinear sensitivity to energy level parameters and mode frequencies.

In this context, we investigate the estimation of the donor and acceptor energies along with the anharmonic mode, which plays a central role in controlling transport characteristics and energy conversion efficiency in molecular systems~\cite{ratner2013, cuevas2017}. Note that such system and system bath parameters have been individually estimated \cite{peng2024enhanced,saleem2023optimal} where the Fischer information was shown to be maximum before the the steadystate is reached \cite{saleem2023optimal}.  The steady-state quantum Fisher information with respect to the donor level can be computed from the density matrix \(\rho(p)\), where \(p\) parametrizes the donor energy. The QFI is defined via the symmetric logarithmic derivative (SLD), which satisfies \(\partial_p \rho = \frac{1}{2}(L \rho + \rho L)\), and is given by \(F_Q(p) = \mathrm{Tr}[\rho L^2]\). When the steady state is diagonal in a fixed basis, which occurs in the absence of coherent superpositions, the QFI reduces to the classical Fisher information (FI) associated with the eigenvalues of \(\rho\), thus satisfying \(F_Q = F_C\). In such cases, classical probability theory yields the optimal estimation strategy, and the quantum limit does not exceed classical bounds~\cite{liu2019}. This equivalence becomes important in systems driven into nonequilibrium steady states by thermal leads, where coherences are suppressed due to strong dephasing or lack of symmetry. 

A central feature of parameter estimation using FI is the dependence of sensitivity on the measurement time. For systems evolving under Markovian dynamics, the FI initially increases quadratically in time due to coherent contributions, but eventually crosses over to a linear scaling dictated by the steady-state structure~\cite{gammelmark2014, albarelli2020}. The optimal time for parameter estimation balances this tradeoff. If measurement duration is too short, sensitivity is limited by quantum projection noise if too long, the estimator variance is bounded by classical stochasticity. For parameters encoded in the steady-state density matrix, the optimal regime corresponds to the long-time limit where QFI per unit time saturates and reflects the system's information content about the parameter.

In donor–acceptor junctions, the donor energy directly controls the energetics of electron transfer and the activation of vibrational modes. Its estimation is relevant for characterizing electron-phonon coupling mechanisms, energy filtering, and the tuning of vibrational sidebands~\cite{haertle2011, schulz2015}. In experimental realizations such as single-molecule transistors based on porphyrin or benzene derivatives, the donor and acceptor moieties correspond to chemically defined groups that are spatially separated and electronically coupled via molecular bridges~\cite{nitzan2003}. The vibrational mode may correspond to torsional or bending motions within the molecule or to localized surface modes when adsorbed on metallic substrates.

The physical parameters that define such systems—donor level position, inter-site coupling strength, vibrational frequency, and vibrational anharmonicity—can be modulated by gate voltages, mechanical strain, or chemical modification. These controls allow access to different dynamical regimes and make parameter estimation experimentally feasible. For instance, scanning tunneling microscopy and transport spectroscopy can access current-voltage characteristics sensitive to level alignment and vibrational resonances~\cite{galperin2006}.

A key insight from quantum estimation theory is that the structure of the Liouvillian spectrum determines the achievable sensitivity~\cite{albarelli2020}. Small Liouvillian gaps suppress sensitivity due to slow relaxation, while large gaps enhance distinguishability between states with nearby parameter values. In the considered rectifier model, the coupling between electronic transitions and an anharmonic mode reshapes the spectrum and thus should modulate the FI landscape. The resulting sensitivity is neither monotonic in coupling strength nor vibrational frequency, as the system may exhibit resonances or dynamical bottlenecks that either enhance or degrade estimation performance\cite{montenegro2023quantum,peng2024dissipative}, precisely serving as the motivation behind the work. Further, FI quantifies potential bounds on the precision achievable through any measurement protocol, motivating its computation and analysis for quantum transport systems. Studies have shown that the QFI can reach a maximum at specific times before decreasing due to decoherence and dissipation effects. Identifying this optimal time is crucial for designing efficient sensing protocols, as measurements performed at this time can achieve the highest precision with minimal resource expenditure \cite{saleem2023optimal,montenegro2023quantum}.

\section{Anharmonic Rectifier Model}
We consider a donor–acceptor molecular junction coupled to an internal vibrational mode that exhibits strong anharmonicity. This vibrational mode, localized within the junction, interacts with the electronic degrees of freedom and may also couple to phononic environments, an aspect discussed later in the manuscript. It can serve as a quantum rectifier \cite{simine2012vibrational,friedman2017effects} since it has an ability to direct current asymmetrically, i.e, allow energy or charge flow more easily in one direction than the other. Unlike a harmonic oscillator with an infinite and evenly spaced spectrum, we truncate the vibrational Hilbert space to its two lowest levels, effectively modeling it as a two-level system (TLS), or equivalently, a spin-½ particle. This constitutes a strong form of anharmonicity, where only the ground and first excited vibrational states are retained. Such a truncation captures the behavior of localized vibrational excitations under strong confinement or rapid relaxation—typical in molecular junctions and quantum dots.

This highly anharmonic vibrational mode interacts asymmetrically with the donor–acceptor system, leading to nonreciprocal transport properties. Specifically, energy or charge flow is directionally dependent: a phenomenon known as rectification. In our model, this effect arises due to three central mechanisms: (i) asymmetric coupling between the vibrational mode and electronic sites—typically the mode modulates either the donor or acceptor, but not both; (ii) truncation to a two-level vibrational subspace, which restricts allowed transitions and breaks spectral symmetry; and (iii) energy-dependent and asymmetric density of states (DOS) in the electronic leads, which can be engineered to favor transport in one direction. Together, these ingredients enable control over the directionality of energy flow, defining what we refer to as a vibrational rectifier.
 The two-level approximation limits the number of accessible excitation channels, introducing a nonlinear energy filtering effect that enhances transition asymmetry. As a result, the system exhibits stronger thermal and electronic rectification than harmonic or weakly anharmonic models. A schematic illustration of the setup is shown in Fig.~1a.
The total Hamiltonian governing the dynamics of the system:
\begin{equation}
\hat{H} = \hat{H}_\mathrm{M} + \hat{H}_\mathrm{L} + \hat{H}_\mathrm{R} + \hat{H}_\mathrm{c} + \hat{H}_\mathrm{vib} + \hat{H}_\mathrm{I}.
\label{eq:Htotal}
\end{equation}

The electronic part of the molecular junction comprises two localized states, representing donor and acceptor orbitals:
\begin{equation}
\hat{H}_\mathrm{M} = \epsilon_d \hat{c}_d^\dagger \hat{c}_d + \epsilon_a \hat{c}_a^\dagger \hat{c}_a,
\end{equation}
where $\hat{c}_d^\dagger$ and $\hat{c}_a^\dagger$ are fermionic creation operators for the donor and acceptor, and $\epsilon_d$, $\epsilon_a$ are their respective onsite energies.

The leads are modeled as non-interacting electron reservoirs:
\begin{equation}
\hat{H}_\mathrm{L} = \sum_{\ell \in \mathrm{L}} \epsilon_\ell \hat{c}_\ell^\dagger \hat{c}_\ell, \quad
\hat{H}_\mathrm{R} = \sum_{r \in \mathrm{R}} \epsilon_r \hat{c}_r^\dagger \hat{c}_r.
\end{equation}

Electron tunneling between the molecule and the leads is described by:
\begin{equation}
\hat{H}_\mathrm{c} = \sum_\ell v_\ell \left( \hat{c}_\ell^\dagger \hat{c}_d + \hat{c}_d^\dagger \hat{c}_\ell \right) + 
\sum_r v_r \left( \hat{c}_r^\dagger \hat{c}_a + \hat{c}_a^\dagger \hat{c}_r \right),
\end{equation}
where $v_\ell$ and $v_r$ denote the tunneling amplitudes to the donor and acceptor sites, respectively.

In the harmonic limit, the internal vibration would be modeled via bosonic operators $\hat{b}_0^\dagger$, $\hat{b}_0$ as:
\begin{equation}
\hat{H}_\mathrm{vib} = \omega_0 \hat{b}_0^\dagger \hat{b}_0,
\end{equation}
with frequency $\omega_0$. 

The interaction between the electronic system and the vibrational mode, limited to vibrationally assisted transitions between donor and acceptor sites, is given by:
\begin{equation}
\hat{H}_\mathrm{I} = \kappa \left( \hat{c}_d^\dagger \hat{c}_a + \hat{c}_a^\dagger \hat{c}_d \right)\hat{\sigma}_x,
\end{equation}
where $\kappa$ is the electron-vibration coupling strength and $\hat{\sigma}_x$ induces transitions between the vibrational states.

In this setup, there is no direct electronic coupling between donor and acceptor orbitals; electronic transitions are mediated entirely by vibrational excitations. This ensures that transport occurs via inelastic tunneling, with the vibrational mode serving as the conduit. Coherent tunneling is thus neglected, a simplification justified by our focus on vibrational dynamics and excitation statistics. While coherent effects may alter current magnitudes, they do not qualitatively change the behavior of the vibrational population—our primary observable.

This vibrationally assisted mechanism is reminiscent of models used to describe current-induced light emission~\cite{galperin2005current,galperin2012molecular}, where electron–hole pair generation couples to an internal mode. One may also interpret the donor and acceptor as HOMO and LUMO orbitals, each potentially coupled to both leads. However, for estimating vibrational observables and establishing fluctuation relations, it is advantageous to exclude coherent contributions. In strongly coupled regimes, each tunneling event is assumed to fully excite or de-excite the TLS—an approximation consistent with our truncated mode description.

This model thus encapsulates key features of vibrational rectification: (i) inelastic, vibrationally mediated transport; (ii) asymmetry in electronic–vibrational coupling; and (iii) strong mode anharmonicity enabling nonlinear filtering. It reproduces expected spin-boson behavior at low temperatures and reveals incoherent, multi-channel excitation dynamics at elevated temperatures.

\subsection{Quantum Master Equation and Rates}
We account for the anharmonicity in the vibrational mode by truncating the harmonic ladder. In this case, we simply use a two-level vibrational Hamiltonian $
\hat{H}_\mathrm{vib} = \frac{\omega_0}{2} \hat{\sigma}_z$
where $\hat{\sigma}_z$ is the Pauli matrix acting in the vibrational subspace. This allows us to proceed in such a way that the dynamics of the system represents that of a two level system through a reduced  density matrix, usually called the spin -fermion mapping \cite{simine2012vibrational}. Using a Born–Markov–Redfield approach, which is appropriate in the weak coupling limit. 
The density vector consists of two populations,
 $p_1(t)$ and $p_2(t)$ representing the occupation probabilities of the donor and acceptor states, respectively.  The reduced density matrix $\hat{\rho}_{} =\{p_1,p_2\}$ obeys the equation
\begin{equation}
\frac{d}{dt} \hat{\rho}_{}(t) = - \int_0^\infty d\tau \, \text{Tr}_{\text{leads}} \left[ \hat{H}_{\text{int}}(t), \left[ \hat{H}_{\text{int}}(t-\tau), \hat{\rho}_{\text{TLS}}(t) \otimes \hat{\rho}_{\text{leads}} \right] \right],
\end{equation}
where $\hat{H}_{\text{int}} = \hat{\sigma}_z \hat{F}_c(t)$ and operators are in the interaction picture. In the Liouville space,
a Pauli master equation follows. Since, the model is well accepted and studies,  for a complete derivation of the master equation and underlying algebra, we simply refer to the following works \cite{simine2012vibrational,friedman2017effects,agarwalla2015full}. The master equation is
\begin{equation}
    \frac{d}{dt} \hat{\rho}(t) = \mathcal{L}\hat{\rho}(t),
\end{equation}  with the superoperator being,
\begin{equation}
\mathcal{L}= \Gamma\begin{pmatrix}
- \alpha_{da}^+ - \alpha_{ad}^+ & \alpha_{da}^- + \alpha_{ad}^- \\
\alpha_{da}^+ + \alpha_{ad}^+ & - \alpha_{da}^- - \alpha_{ad}^- 
\end{pmatrix}.
\end{equation}
where $
\alpha_{da}^+= f_L [1 - f_R^{+}] 
\alpha_{ad}^+ = f_R [1 - f_L^{+}] 
\alpha_{ad}^- = f_R [1 - f_L^{-}] 
\alpha_{da}^- = f_L [1 - f_R^{-}]. 
$
The Fermi distributions and the modified Fermi distributions are $f_L= (\exp\left(\frac{e_d - \mu_L}{k_B T_L}\right) + 1)^{-1},  
f_R = (\exp\left(\frac{e_a - \mu_L}{k_B T_R}\right) + 1)^{-1}, 
f_{L}^{\pm} = (\exp\left(\frac{e_d \pm \omega_0 - \mu_L}{k_B T_L}\right) + 1)^{-1}, 
f_{R}^{\pm} = (\exp\left(\frac{e_a \pm \omega_0 - \mu_R}{k_B T_R}\right) + 1)^{-1}
$. $
\alpha_{da}^+ $
describes an inelastic tunneling process in which an electron transfers from the donor site (coupled to the left lead) to the acceptor site (coupled to the right lead) while exciting the vibrational mode. This process requires that the left lead has an occupied state at energy $\epsilon_d$, and the right lead has an unoccupied state at energy $\epsilon_a + \omega_0$, corresponding to the absorption of vibrational energy $\omega_0$ during the electron transfer. $
\alpha_{ad}^+ $
corresponds to the reverse process, where an electron moves from the acceptor to the donor site while exciting the vibrational mode. This occurs when the right lead is occupied at energy $\epsilon_a$, and the left lead has an empty state at $\epsilon_d + \omega_0$, allowing energy $\omega_0$ to be absorbed by the mode. $
\alpha_{da}^- $
captures a forward tunneling process accompanied by vibrational de-excitation, in which an electron transitions from the donor to the acceptor site while emitting energy $\omega_0$ into the vibrational mode. This process is enabled when the left lead is occupied at $\epsilon_d$ and the right lead has an available state at $\epsilon_a - \omega_0$. $
\alpha_{ad}^- $
represents the reverse tunneling process with vibrational de-excitation, where an electron moves from the acceptor to the donor, releasing energy $\omega_0$ into the vibrational mode. The right lead supplies the electron at $\epsilon_a$, and the left lead accepts it at $\epsilon_d - \omega_0$.
The general time-dependent solution is
\begin{align}
\label{eq-pop}
    p_1(t) &= p_1^{\text{ss}} + (p_1(0) - p_1^{\text{ss}}) e^{-\gamma t} \\
    p_2(t) &= p_2^{\text{ss}} + (p_2(0) - p_2^{\text{ss}}) e^{-\gamma t},
\end{align}
where $\gamma = \alpha_{da}^+ + \alpha_{ad}^+ + \alpha_{da}^- + \alpha_{ad}^-$ is the total decay rate and $p_i^{\text{ss}}$ are steady-state values. For our numerical analysis, we take the example of a
charge-transfer complex, the anthracene-PMDA system, to serve as prototype in estimating Fischer Information pertaining to target parameters. In this system, the donor is anthracene with an energy of HOMO $\approx -5.4$ eV. The acceptor is PMDA whose energy of the LUMO $\approx -3.8$ eV. The exciton binding energy $\sim 0.7$ eV. Its major vibrational modes are 
     C=C Stretching frequency $\sim 1580$ cm$^{-1}$ ($\sim 0.196$ eV),
         C--C--H Bending mode $\sim 1120$ cm$^{-1}$ ($\sim 0.139$ eV) and  C--C--C bending: $\sim 740$ cm$^{-1}$ ($\sim 0.092$ eV)\cite{fonari10.1063/1.4936965}. We separately consider each vibration to represent the anharmonic mode of the model Hamiltonian in order to evaluate the Fischer information. In Fig. (1b), we show the time evolution of the probabilities of the two states evaluated at the three cases where each case represent the different anharmonic modes. The solid lines represent $p_1(1)$ while the dotted lines represent $p_2(t)$. Not much difference is seen on the dynamics. However, the Fischer Information behaves differently for each anharmonic mode wich we discuss next. 
\section{Results and Discussion on Fischer Information}
Due to decoupled populations and coherences, to estimate any parameter $\theta$ of the system, eg. $\epsilon_d$ or $\omega_0$ or $\Gamma$ etc, we can simply use the classical Fisher information at time $t$ instead of the quantum Fischer Information. It is defined as
\begin{equation}
\label{eq-I}
I(\theta) = \sum_i \frac{1}{p_i(t)} \left( \frac{\partial p_i(t)}{\partial \theta} \right)^2
\end{equation}
representing the time-dependent sensitivity of the population distribution to the  target parameter.

\begin{figure}
\centering
\includegraphics[width = 16cm]{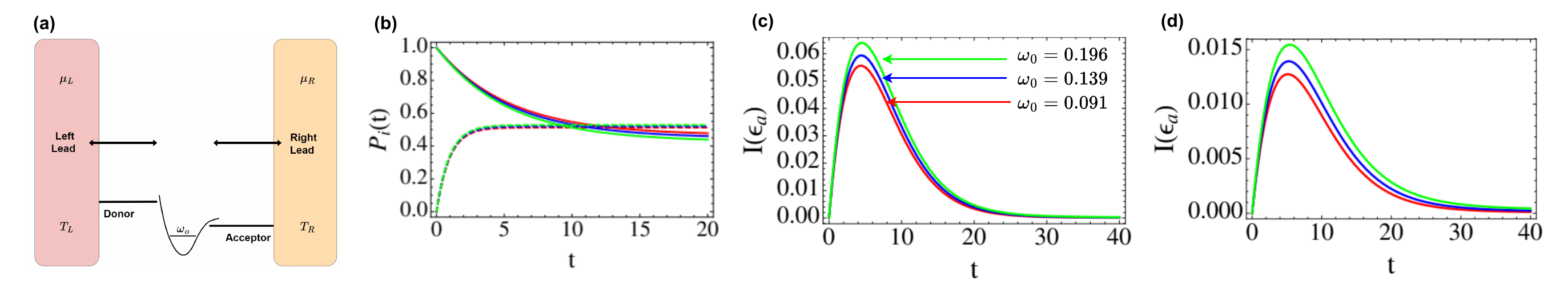}
\caption{ (a) Schematic representation of the donor-acceptor rectifier coupled to an anharmonic mode. (b) Time evolution of the occupation probablities of the donor and acceptor states for three different vibrational energies, $\omega_0=0.019 eV, 0.139 eV$ and $0.196 eV$. The other system parameters are fixed as follows: $\epsilon_d=-5.4eV, \epsilon_a=-3.8eV, \mu_L =1eV, \mu_R=-1eV, T_L = 2eV$ and $T_R=1eV$ (with $\hbar = 1$ and $k_B=1$).  Time-dependent Fischer Information for estimating acceptor state's energy, $I(\epsilon_a)$ for the three different vibrations at (c) $\mu_L=1eV$ and $\mu_R =-1eV$ and (d) low bias regime, $\mu_L=0.1eV$ and $\mu_R =-0.1eV$. All energy and temperatures are in eV. Time is in the units of $\Gamma$ which is set to be 0.7eV.}
\label{fig-1}
\end{figure}

Firstly, we start by calculating the Fischer Information for the energy of the acceptor state which we represent as $I(\epsilon_a)$. $I(\epsilon_a)$ is numerically calculated using the definitions of the time-dependent solutions of the populations, Eq. (\ref{eq-pop} and 11) on Eq. (\ref{eq-I}) with $\theta = \epsilon_a$. We evaluate the quantity separately for three cases: $\omega_0=0.091eV, 0.139eV$ and $0.196eV$. Each of the three different vibrational modes are depicted as a separate curve in the figures and represent a a separate calculation. Note that, such a choice of system parameter to be estimated is existent in the domain of open quantum systems\ \cite{saleem2023optimal}. In Fig. (1c), we numerically evaluate $I(\epsilon_a)$ as a function of time for three different anharmonic frequencies. All of the three curves show the typical optimal behavior as reported earlier \cite{saleem2023optimal} re-establishing the fact that, the Fischer Information is maximum before the system reaches a steadystate. What is interesting to observe is the fact that, higher the anharmonic mode's energy, higher is the value of the information. This is an indication that, vibrations with higher energies tend to increase the Fischer Information. When we decrease the bias between  the two terminals (low bias regime, achieved by tuning the chemical potentials of the left and right leads), the magnitude of $I(\epsilon_a)$ decreases drastically, as evident from Fig. (1d). The effect of the vibration mode remains the same. Upon increasing the bias to a large extent (high bias limit), $I(\epsilon_a)$ for all the modes is identical as seen from Fig. (2d) with a larger value of the information in comparison to lower differences between the chemical potentials. Another subtle observation is the shifting of the optimal time of $I(\epsilon_a)$ to earlier times as the bias keeps increasing. This can confirmed from the peak positions of Fig. (1d,c) and Fig. (2a) which shifts to earlier times when $\mu_L-\mu_R =6eV, 2eV$ and $0.2eV$ respectively.  We conclude by claiming that $\epsilon_a$, a system parameter can be estimated better if the bias is high, even in the presence of vibronic coupling of any energy. However, this comes at the cost of measuring the system at  earlier times. As the bias decreases, the optimal time for measurement happens at larger times, and vibrations with higher energies have a positive correlation with the estimation. In Fig. (2 b, c and d), we choose the donor's energy for estimation and calculate $I(\epsilon_d)$, where also we observe the existence of the reported optimal time of estimation. As can be seen from the identical nature of the curves in each figure, the vibrations have no significant effect on  $I(\epsilon_d)$. Again, the optimal time of estimation shifts to earlier times as the bias increases, similar to the the observation in $I(\epsilon_a)$. A point to note is that $I(\epsilon_a)\gg I(\epsilon_d)$ as can be confirmed by comparing the magnitudes of the Fischer Information in Fig. (1c,d) and Fig. (2a) with Fig. (2b,c and d). Thus, the estimation of acceptor energy is always better than the donor's energy.

\begin{figure}
\centering
\includegraphics[width = 17cm]{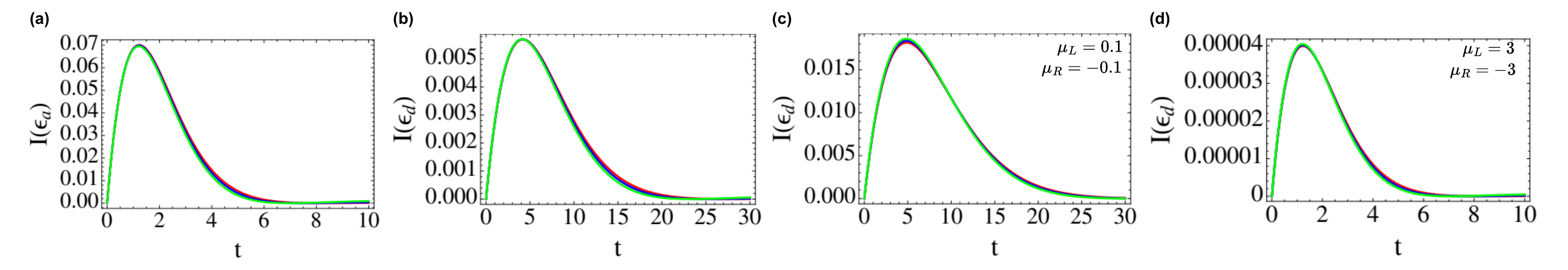}
\caption{ (a) Time-dependent Fischer Information for estimating acceptor state's energy, $I(\epsilon_a)$ for the three different vibrations at high bias, $\mu_L=3eV$ and $\mu_R =-3eV$. Time-dependent Fischer Information for donor energy, $I(\epsilon_d)$ at (b) mid-bias
(c) low bias and (d) high bias. }
\label{fig2}
\end{figure}

\begin{figure}
\centering
\includegraphics[width = 17cm]{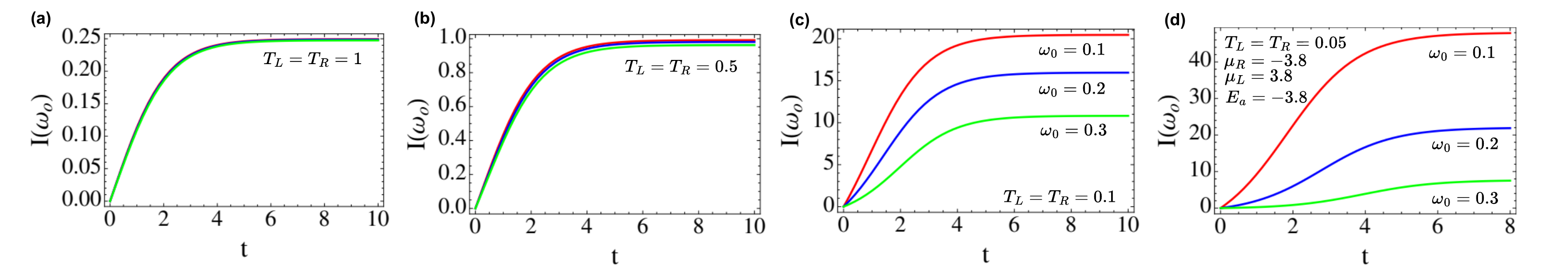}
\caption{ Time-dependent Fischer Information for estimating the vibration mode's energy, $I(\omega_0)$ for the three different vibrations at different temperatures, (a) $T_L=1 =T_R$ (b) $T_L =T_R = 0.5$ (c) $T_L =T_R = 0.1$ and (d) $T_L =T_R = 0.05$ at a high bias of $\mu_L -\mu_R = 7.6$. Note the loss of optimal time of estimation to a saturating nature.
}
\label{fig3}
\end{figure}

We now move to estimating the anharmonic mode's energy under different conditions by evaluating $I(\omega_0)$ under different system conditions. The results are shown in Fig. (3a-d). The time-dependent behaviour of $I(\omega_0)$ is totally different from $I(\epsilon_a)$ and $I(\epsilon_d)$. It shows a saturating behaviour instead of an optimal time for estimation. The saturating behavior is conclusive of the fact that, the estimation of the anharmonic mode is best once the system reaches a steadystate. We observe this to be true for any system parameter we vary and any anharmonic mode we chose. This in stark contrast to existing reports which claimed that system parameters have an optimal time of measurement before the system reaches a steadystate.  We show a few numerical results for different parameters. 

\begin{figure}
\centering
\includegraphics[width = 17cm]{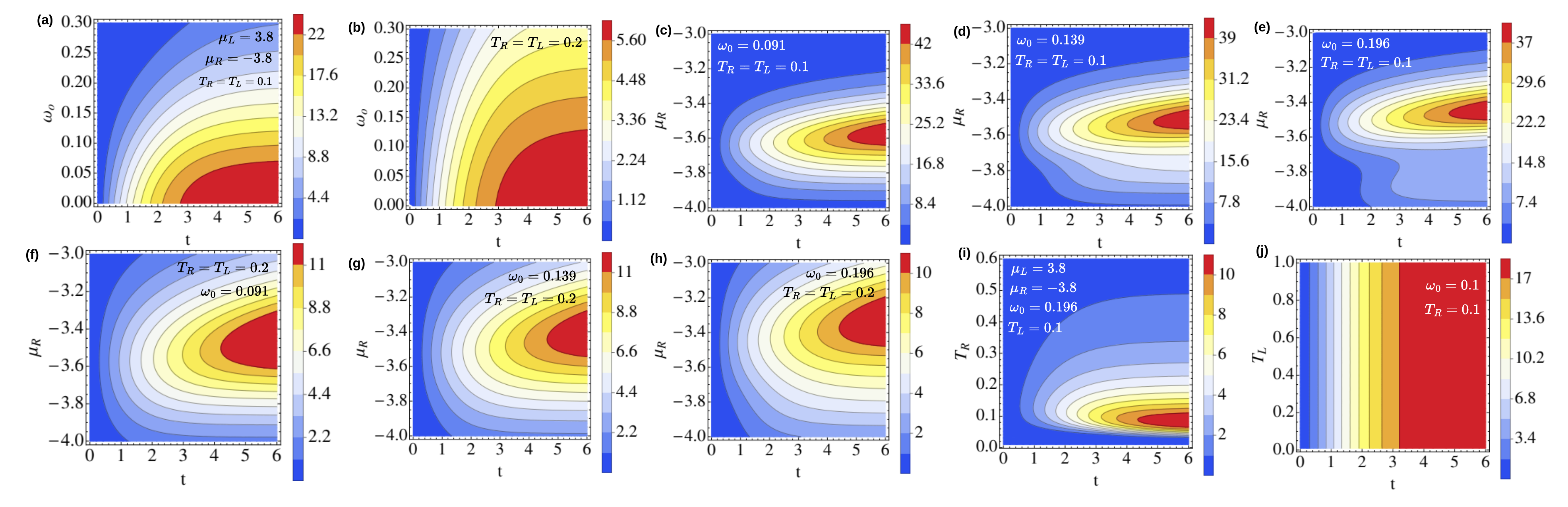}
\caption{ Contours of time-dependent $I(\omega_0)$ when another system parameter is varied evaluated at high bias. 
}
\label{fig4}
\end{figure}

In Fig. (3a), we consider a high bias case ($\mu_L=3.8eV, \mu_R =-3.8eV$) evaluated at equal temperatures. In this case, we donot observe any effect of the magnitude of the anharmonic mode's energy to influence the Fischer Information. All the anharmonic modes show a saturating behavior. As we decrease the temperature, the magnitude of $I(\omega_0)$ increases fourfold, Fig. (3b) and slight differences in magnitude starts showing up for the three different anharmonic modes. As we decrease the temperature further, there is a dramatic increase in $I(\omega_0)$ and the effect of vibrations become clearer. A lower magnitude of vibrational energy leads to a larger $I(\omega_0)$ as evident from Fig.(3c and d). We conclude by saying that estimation of a vibrational mode's energy is better at lower temperatures with lower vibrational energies leading to higher probability of estimation.

To analyse the role of other bath parameters on $I(\omega_0)$, we numerically create two-dimensional data as a function of time and express the data as contours in Fig. (4a-j). In all the parameter variations, steadystate is the best time to estimate the anharmonic mode's energy. We now systematically explain the observations from the contour plots one by one. In Fig. (4a), we vary the frequency of the anharmonic mode at all times equal temperatures. The information is highest at $\omega_0=0$ and at the steadystate and nonlinearly decreases along the $\omega_0$ axis. We increase the temperature under the same conditions and observe that the magnitude decreases. We then chose a frequency $\omega_0=0.091eV$ and vary the chemical potential of the left lead at equal temperatures, Fig. (4c). We see a maximum at $\mu_R = -3.509eV$ indicative of the resonance $ \epsilon_a+\omega_0 = \mu_R$. The Fermi functions leading to this resonance is present in the rate $\alpha_{da}^+$. Hence, the amount of information is dictated by the inelastic electron transfer process from the donor to the acceptor complex whilst exciting the vibration mode, i.e forward transfer. This is true since, high bias mostly favours the forward transfer. The maximum shifts upward in the $\mu_L$ axis as we change to $\omega_0=0.139 eV$ (Fig. 4d) and $\omega_0=0.196eV$ (Fig. 4e). All the maxima are at the mentioned resonances. The magnitude of $I(\omega_0)$ is also seen to decrease as the anharmonic frequency increases in these three figures. As we increase the temperature, the magnitude of $I(\omega_0)$ decreases (Fig.4f,g and h) in comparison to the earlier case of lower temperature (Fig. 4 c,d and e). Also the effect of the vibrational mode on the intensity of information starts to be less visible. In the high-bias limit, changing the temperature gradient alters the nature of the behaviour of $I(\omega_0)$ in comparison to the previous cases. Although the optimal time for estimation is at the steadystate, the largest value of the information is when $T_L\approx T_R$ as seen from the location of the maxima in Fig. (4i). The Fischer Information decreases as the temperature of left lead is lowered or increased for a fixed $T_R$. Further, when $T_R$ is fixed, varying $T_L$ doesn't  have any effect on the Fischer Information, $I(\omega_0)$ as evident from the flat vertical contours in Fig. (4j) . 
\begin{figure}
\centering
\includegraphics[width = 16.5cm]{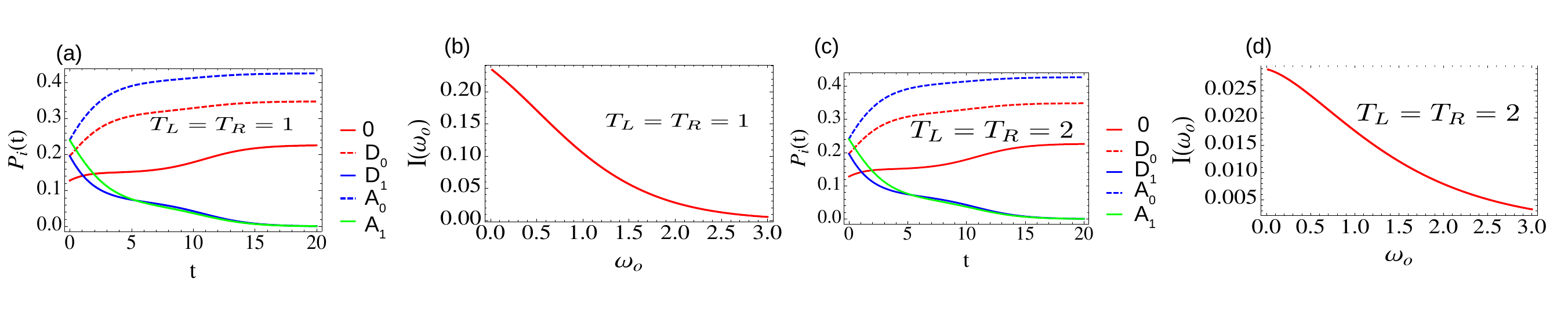}
\caption{ (a, c) Population dynamics of the rectifier with bosonic states included in the density matrix evaluated at two different temperatures in the high bias limit. (b,d) Steadystate Fischer Information, $I(\omega_0)$ as a function of $\omega_0$ for the same parameters as before. $\gamma_0=0.5eV, \Gamma_R=\Gamma_L =1eV.$ The phonon bath temperature is assumed to be the same as the lead temperature.
}
\label{fig4}
\end{figure}

\section{Steadystate Fischer Information with a Larger Liouville Space}

We extend our analysis to include a larger basis set by including the vibrational mode to states $|n=0,1\rangle$ and assume single occupancy of the donor and acceptor. The larger basis states that we consider is composed of the five states
\(
|0\rangle, |D_0\rangle, |D_1\rangle, |A_0\rangle, |A_1\rangle
\). $|0\rangle$ is an empty state,$ |D_0\rangle$ is a donor occupied with  ground state vibration, $|D_1\rangle$ is a donor occupied, with an excited vibration, 
$|A_0\rangle$ is an acceptor occupied with a ground state vibration  and $|A_1\rangle$ is an acceptor occupied with an excited vibration. 
In this basis, assuming the standard Born-Markov Approximations coupled with the rotating wave approximation, the superoperator takes the form

\begin{equation}
\mathcal L=\begin{pmatrix}
-\bigl(\Gamma_{L0}^++\Gamma_{L1}^++\Gamma_{R0}^++\Gamma_{R1}^+\bigr)
&\Gamma_{L0}^- &\Gamma_{L1}^- &\Gamma_{R0}^- &\Gamma_{R1}^- \\
\Gamma_{L0}^+
&-\bigl(\Gamma_{L0}^-+k_0^{DA}+\gamma_\uparrow\bigr)
&\gamma_\downarrow &k_0^{AD} &0\\
\Gamma_{L1}^+
&\gamma_\uparrow
&-\bigl(\Gamma_{L1}^-+k_1^{DA}+\gamma_\downarrow\bigr)
&0 &k_1^{AD}\\
\Gamma_{R0}^+
&k_0^{DA} &0 &-\bigl(\Gamma_{R0}^-+k_0^{AD}+\gamma_\uparrow\bigr)
&\gamma_\downarrow\\
\Gamma_{R1}^+
&0 &k_1^{DA} &\gamma_\uparrow
&-\bigl(\Gamma_{R1}^-+k_1^{AD}+\gamma_\downarrow\bigr)
\end{pmatrix}.
\end{equation}
with 
 $|0\rangle\leftrightarrow|D_0\rangle$ transition involving energies $\epsilon_d$ and rates $\Gamma_L^\pm(\epsilon_d)$,  $|0\rangle\leftrightarrow|D_1\rangle$ involving $\epsilon_d+\omega_0$ and rates $\Gamma_L^\pm(\epsilon_d+\omega_0)$, 
transition $|0\rangle\leftrightarrow|A_0\rangle$ at $\epsilon_a$ involving rates $\Gamma_R^\pm(\epsilon_a)$ and 
transition $|0\rangle\leftrightarrow|A_1\rangle$ at $\epsilon_A+\omega_0$ involving rates $\Gamma_R^\pm(\epsilon_A+\omega_0)$.
The  vibronic donor to acceptor transitions $|D_0\rangle\leftrightarrow|A_0\rangle$ and  $|D_1\rangle\leftrightarrow|A_1\rangle$ involve the rates $k_0^{DA/AD},k_1^{DA/AD}$ while the vibrational relaxation on the donor  and acceptor, $|D_1\rangle\leftrightarrow|D_0\rangle$ and $|A_1\rangle\leftrightarrow|A_0\rangle$ are $\gamma_{\uparrow,\downarrow}$. 
The donor-lead tunneling rates are
\begin{equation}
\Gamma_{L n}^+ = \Gamma_L F_n\, f_L(\varepsilon_d + n\omega_0), \qquad
\Gamma_{L n}^- = \Gamma_L F_n\, \bigl[1 - f_L(\varepsilon_d + n\omega_0)\bigr].
\end{equation} \(\Gamma_L\) is the hybridization with the left lead, and \(F_n = \abs{\langle n | \hat{X} | 0 \rangle}^2\) is the Franck--Condon factor with \(\hat{X} = \exp\left[\frac{\lambda}{\omega_0}(b^\dagger - b)\right]\). The acceptor--lead tunneling rates are
\begin{equation}
\Gamma_{R n}^+ = \Gamma_R \delta_{n0} f_R(\varepsilon_a), \qquad
\Gamma_{R n}^- = \Gamma_R \delta_{n0} \bigl[1 - f_R(\varepsilon_a)\bigr],
\end{equation}
with \(f_R(\varepsilon)\) being the Fermi function for the right lead and \(\Gamma_R\) the tunneling amplitude. Donor--acceptor inelastic transfer rates mediated by phonons are given by
\begin{equation}
k_n^{DA} = \Gamma_{DA} F_n \left[1 + n_B(\omega_0)\right], \qquad
k_n^{AD} = \Gamma_{AD} F_n\, n_B(\omega_0),
\end{equation}
where \(n_B(\omega_0) = \frac{1}{e^{\omega_0/T_v} - 1}\) is the Bose distribution function corresponding to the vibrational bath.
Vibrational relaxation and excitation rates due to the thermal phonon environment are $
\gamma_\uparrow = \gamma_0 n_B(\omega_0), \qquad
\gamma_\downarrow = \gamma_0 \left[1 + n_B(\omega_0)\right].
$
Physically, the terms \(\Gamma_{L n}^{+}\) and \(\Gamma_{R n}^{+}\) describe injection of electrons from the leads into the donor or acceptor site, with \(\Gamma_{L n}^{+}\) allowing for vibronic excitation during tunneling into the donor. The terms \(\Gamma_{L n}^{-}\) and \(\Gamma_{R n}^{-}\) correspond to electron extraction processes. The donor--acceptor rates \(k_n^{DA}\) and \(k_n^{AD}\) incorporate the effects of vibrational excitation and relaxation on electron hopping between the donor and acceptor. The coefficients \(\gamma_\uparrow\) and \(\gamma_\downarrow\) describe pure phonon-driven transitions among vibrational levels, independent of electronic motion, allowing the system to thermalize within vibrational subspaces. All rates are temperature- and bias-dependent, determining the directionality and magnitude of steady-state currents and rectification. We assume equal coupling and neglect the Frank-Condon factors since the space is spanned by a single photon. We can solve for the steadystates values of the five populations using $\dot\rho={\cal L}\rho =0$, where $\rho$ is the density vector containing the five populations. The time evolution of the five populations for two different temperatures are shown in Fig. (5a and c). Using the steadystate values, evaluate the Fischer Information using Eq. (\ref{eq-I}) with $\theta=\omega_0$. In Fig. (5b and c), we show $I(\omega_0)$ as a function of $\omega_0$ for the same numerical parameters as before for two different temperatures in the high bias limit. As $\omega_0$ increases the information nonlinearly decays to zero for both the cases. Lower temperatures lead to a larger values of the Fischer Information. The information decays to zero since the exceptionally high vibrational energies are incapable of being accessed by the system to facilitate any electron transfer between the leads leading to loss of estimation. 

\section{Conclusion}

In summary, we theoretically  studied a nonequilibrium donor-acceptor quantum junction rectifier system  coupled to an anharmonic vibrational mode, by treating the vibrational subspace as a two-level system and well as a five level system with an inclusive phonon mode. By analyzing  Fischer information for key parameters (donor energy $\varepsilon_d$, acceptor energy $\varepsilon_a$, and vibrational frequency $\omega_0$), we uncovered distinct estimation behaviors under various bias and temperature conditions. The Fischer information for the acceptor energy, $I(\varepsilon_a)$, exhibits a pronounced maximum at an optimal interrogation time and grows significantly with higher vibrational frequency and larger bias voltage, indicating that $\varepsilon_a$ can be estimated most precisely under strong vibronic coupling and far-from-equilibrium. In contrast, the Fischer information for the donor energy, $I(\varepsilon_d)$, remains consistently lower than $I(\varepsilon_a)$ and is comparatively insensitive to the vibrational frequency, reflecting the intrinsically weaker imprint of $\varepsilon_d$ on the system’s nonequilibrium dynamics. Meanwhile, the information associated with the vibrational mode frequency, $I(\omega_0)$, does not show a distinct peak in time but instead saturates as the system evolves; this $I(\omega_0)$ is enhanced at low temperatures and for lower vibrational frequencies, suggesting that reduced thermal noise and strong anharmonicity improve the distinguishability of the vibrational frequency parameter. These findings illustrate how bias-driven transport and strong mode anharmonicity confer parameter-specific Fisher information profiles: acceptor-level estimation is most favorable and time-dependent, donor-level estimation is intrinsically limited, and vibrational frequency estimation requires low-noise conditions and exhibits a long-time information plateau. The pronounced $I(\varepsilon_a)$ suggests that time-resolved transport or spectroscopic measurements in a molecular junction (under high bias and involving high-frequency vibrational modes) could be optimized to accurately determine acceptor site energies. The weaker and less sensitive $I(\varepsilon_d)$ indicates that donor-level energies would be more challenging to extract and may require complementary techniques or longer integration times. Furthermore, the strong temperature dependence of $I(\omega_0)$ implies that vibrational mode frequencies in nanoscale devices can be more precisely inferred in low-temperature environments, where thermal fluctuations are suppressed. Overall, this work demonstrates the value of Fisher information analysis in elucidating how different system parameters manifest in nonequilibrium observables. Such insight provides a guiding framework for designing metrological protocols and sensing schemes in molecular-scale devices, where optimizing bias conditions, vibrational coupling, and timing can significantly enhance the precision of parameter estimation.

\bibliographystyle{elsarticle-num} 
 \bibliography{ref.bib}

\end{document}